\documentclass{aa}  

\usepackage{graphicx}
\graphicspath{{./images/}}
\usepackage{txfonts}
\usepackage[colorlinks]{hyperref}
\hypersetup{
	linkcolor = blue,
	anchorcolor = blue,
	citecolor = blue,
	filecolor = blue,
	urlcolor = blue
}
\usepackage[T1]{fontenc}
\usepackage{amsmath}
\usepackage[numbered]{bookmark}
\usepackage{textgreek}
\usepackage{upgreek}
\usepackage{multirow}

\begin{document}

	\title{Optimized sampling of SDSS-IV MaStar spectra for stellar classification using supervised models}
	
	
	\author{R. I. El-Kholy\thanks{E-mail: \href{mailto:relkholy@sci.cu.edu.eg}{relkholy@sci.cu.edu.eg} (RIE)}
		\and
		Z. M. Hayman
	}
	
	\institute{Department of Astronomy, Space Science, and Meteorology, Faculty of Science, Cairo University, Giza 12613, Egypt
	}
	
	\date{Received XXX 00, 0000; accepted YYY 00, 0000}
	
	
	\abstract
	{Supervised machine learning models
		are increasingly being used for solving the problem of stellar classification of spectroscopic data. However, training such models requires a large number of labelled instances, the collection of which is usually costly in both time and expertise.}
	{Active learning algorithms minimize
		training dataset sizes by keeping only the most informative instances. This paper explores the application of active learning to sampling stellar spectra using data from a highly class-imbalanced dataset.}
	{We utilize the MaStar library from
		the SDSS DR17 along with its associated stellar parameter catalogue. A preprocessing pipeline that includes feature selection, scaling, and dimensionality reduction is applied to the data. Using different active learning algorithms, we iteratively query instances, where the model or committee of models exhibits the highest uncertainty or disagreement, respectively. We assess the effectiveness of the sampling techniques by comparing several performance metrics of supervised-learning models trained on the queried samples with randomly-sampled counterparts. Evaluation metrics include specificity, sensitivity, and the area under the curve; in addition to the Matthew's correlation coefficient, which accounts for class imbalance. We apply this procedure to effective temperature, surface gravity, and iron metallicity, separately.}
	{Our results demonstrate the
		effectiveness of active learning algorithms in selecting samples that produce performance metrics superior to random sampling and even stratified samples, with fewer training instances.}
	{Active learning is recommended for 
		prioritizing instance labelling of astronomical-survey data by experts or crowdsourcing to mitigate the high time cost. Its effectiveness can be further exploited in selection of targets for follow-up observations in automated astronomical surveys.}
	
	
	\titlerunning{Active learning for stellar spectral classification}
	\authorrunning{R. I. El-Kholy \& Z. M. Hayman}
	\maketitle
	%
	
	\section{Introduction}
	Stellar spectra can be divided into seven main spectral classes according to the Harvard scheme of stellar spectral classification. These classes; O, B, A, F, G, K, and M; follow a sequence represented by the effective temperature of stellar atmospheres, where the hottest stars belong to class O ($T_{\text{eff}} \gtrsim 25,000$ K) and the coolest belong to class M ($2,000$ K $< T_{\text{eff}} < 3,500$ K). Each of these main classes can be further divided into 10 subclasses from 0 to 9, where 0 represents the hottest stars within that class and 9 the coolest. Morgan and Keenan later proposed appending a luminosity class (Ia, Ib, II, III, IV, and V) to the main class and subclass (e.g. our Sun is of class G2V). The luminosity class depends on the surface gravity of stars, often represented as $\log{g}$ in stellar parameters catalogues, where luminous supergiants with the least $\log{g}$ values belong to class 'Ia` and dwarfs with the largest values belong to class 'V`. The modified system became known as the MK classification system. A review of stellar spectral classification can be found in \citet{Giridhar2010}.
	
	Stellar spectral classification of large numbers of stars is essential to studies of stellar populations and galactic formation history. In the past, stellar spectral classification has been done by human experts. They had to visually inspect each of the spectra to do so. With the advancement of computational capabilities and the introduction of machine learning (ML) algorithms, more sophisticated techniques have been applied to classify stellar spectra. Among those are $\chi^2$-minimization, artificial neural networks (ANN), and principal component analysis (PCA) \citep[e.g.][]{Gulati1994, Singh1998, BailerJones1998, Manteiga2009, Gray2014, Kesseli2017, Fabbro2017}.  With the avalanche of stellar spectroscopy data pouring from telescope surveys, the use of ML algorithms has been increasing and has proven capable of reducing the error and improving the accuracy of stellar spectral classification \citep{Sharma2019}. However, for any supervised ML algorithm to be applied to the stellar classification problem, a large sample of labelled data has to be collected and curated for the training of the model, which is very costly in terms of both time and expertise. This has always been a limitation for the use of supervised ML techniques, and is especially prominent for applications of deep learning (DL) frameworks. Attempts to tackle this problem by crowdsourcing the classification have been made, such as in the case of the Galaxy Zoo Project \citep{Lintott2010} which eventually started to include many other applications\footnote{\url{https://www.zooniverse.org/}}, and have indeed been effective to some extent. However, this approach in itself suffers from two limitations: (i) For certain tasks, many non-expert volunteers become uncertain of their answers, which might lead to inaccurate labelling which would eventually reflect in the poor performance of models trained using that data; and (ii) the crowdsourcing process does not resolve the problem of the time-cost completely. Some efforts have been employed to solving the first limitation, by careful curation of the questions and taking the confidence level of the volunteers into account, with some success \citep{Song2018}. However, this can also exacerbate the time-cost issue. Another more-effective approach that can minimize the size of the required training dataset while keeping fewer high-quality instances for labelling is active learning (AL) \citep{Lughofer2012}. The use of AL algorithms has been shown to give favourable results in many astronomical applications such as stellar population studies, photometric supernova classification, galactic morphology, and anomaly detection for time-domain discoveries \citep[e.g.][]{Solorio2005,Richards2011,Ishida2018,Walmsley2019,Ishida2021}.
	
	In this work we apply AL algorithms to a set of stellar spectra to study the efficiency of the sampling techniques in selecting instances that are informative and representative of the overall distribution of the data pool and investigate whether the performance of models trained using the selected instances is comparable to that of models trained on randomly-sampled instances, or even stratified samples. We use the MaNGA Stellar Library (MaStar) \citep{Yan2019}, which is highly imbalanced, from the seventeenth data release (DR17) of the Sloan Digital Sky Surveys (SDSS) \citep{Abdurrouf2022}. We start by applying a preprocessing pipeline to the data. We use random sampling to establish a baseline for comparison. We vary both the initial batch size and the number of additional instances sampled using each algorithm. Supervised ML algorithms are then trained using each sample and their performances on test sets are compared. Several metrics are applied for comparing the performances. The process is implemented for three stellar parameters: effective temperature, surface gravity (in terms of $\log{g}$), and iron metallicity. We finally test the progression of the performance of spectral classification with the increase in the number of selected instances, and demonstrate how AL sampling produces results superior to both random and stratified sampling even with less than half the sample size.
	
	The paper is structured as follows. In Section \ref{sec:data}, we give an overview of the spectral dataset used in this work. In Section \ref{sec:method}, we describe the preprocessing steps applied to the data, illustrate the AL algorithms employed, briefly illustrate each of the supervised learning models used for classification, and define the set of performance metrics used for model assessment. In Section \ref{sec:results}, we present our results and discuss their potential interpretations. Finally, in Section \ref{sec:conclusions}, the summary and conclusions of this study are provided.
	
	\section{Data}
	\label{sec:data}
	In this work, we use the final version of the MaStar library from the SDSS DR17 \citep{Abdurrouf2022}. MaStar is a large library of high-quality calibration empirical stellar library. The MaStar data are obtained using the Baryon Oscillation Spectroscopic Survey (BOSS) spectrograph \citep{Smee2013,Drory2015}, the same as the main MaNGA survey, mounted on the Apache Point Observatory 2.5m telescope \citep{Gunn2006}. The same fibre system used by the MaNGA survey was used as well. The targets of the MaStar library were chosen to cover a wide range of parameter space. The MaStar spectra cover a wavelength range of $3,622 - 10,354 \,\r{A}$ with a spectral resolution of $R \sim 1,800$. The first release of MaStar is presented in \citet{Yan2019} and its final version will be detailed in Yan et al. (in preparation).
	
	Empirical spectra are obtained by observing real stars. Hence, they are not subject to many of the limitations of synthetic spectra produced by theoretical models \citep{Kurucz2011,Dupree2016}. However, empirical spectral libraries are limited by the wavelength range, spectral resolution, and parameter-space coverage. This makes the high quality and wide coverage of the MaStar library particularly optimal for use in data-driven experiments. In this work, we only use the good-quality visit spectra included in the \texttt{mastar\_goodspec} file\footnote{\url{https://data.sdss.org/sas/dr17/manga/spectro/mastar/v3_1_1/v1_7_7/mastar-goodspec-v3_1_1-v1_7_7.fits.gz}}. This library has a flux calibration accurate to 4\% \citep{Imig2022}. SDSS DR17 also includes a Value Added Catalogue (VAC)\footnote{\url{https://data.sdss.org/sas/dr17/manga/spectro/mastar/v3_1_1/v1_7_7/vac/parameters/v2/mastar-goodstars-v3_1_1-v1_7_7-params-v2.fits}} containing four sets of different stellar parameter measurements. Each measurement uses different methods, the details of which can be found in the respective papers (\citet{Hill2021,Imig2022}; Chen et al. (in preparation); Lazarz et al. (in press); and Hill et al. (in press)); and a detailed comparison will be presented in Yan et al. (in preparation). The same VAC also includes the median values of these methods when available and robust, along with the uncertainties of these medians based on the quality assessments of each set of measurements. We rely on these median columns in our current work.
	
\begin{figure}
		\centering
		\includegraphics[width=\columnwidth, trim= 9mm 0mm 9mm 0mm, clip]{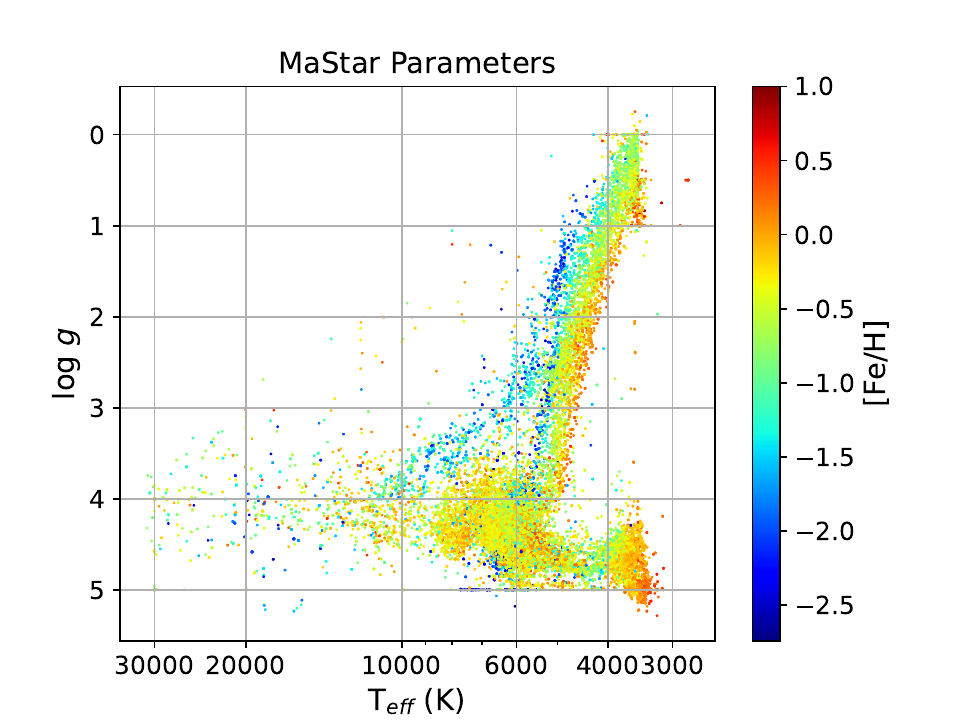}
		\caption{The MaStar library stellar parameter distribution after final quality cuts, colour-coded by iron metallicity.}
		\label{fig:mastar_params}
	\end{figure}

\begin{figure}
		\centering
		\includegraphics[width=0.75\columnwidth]{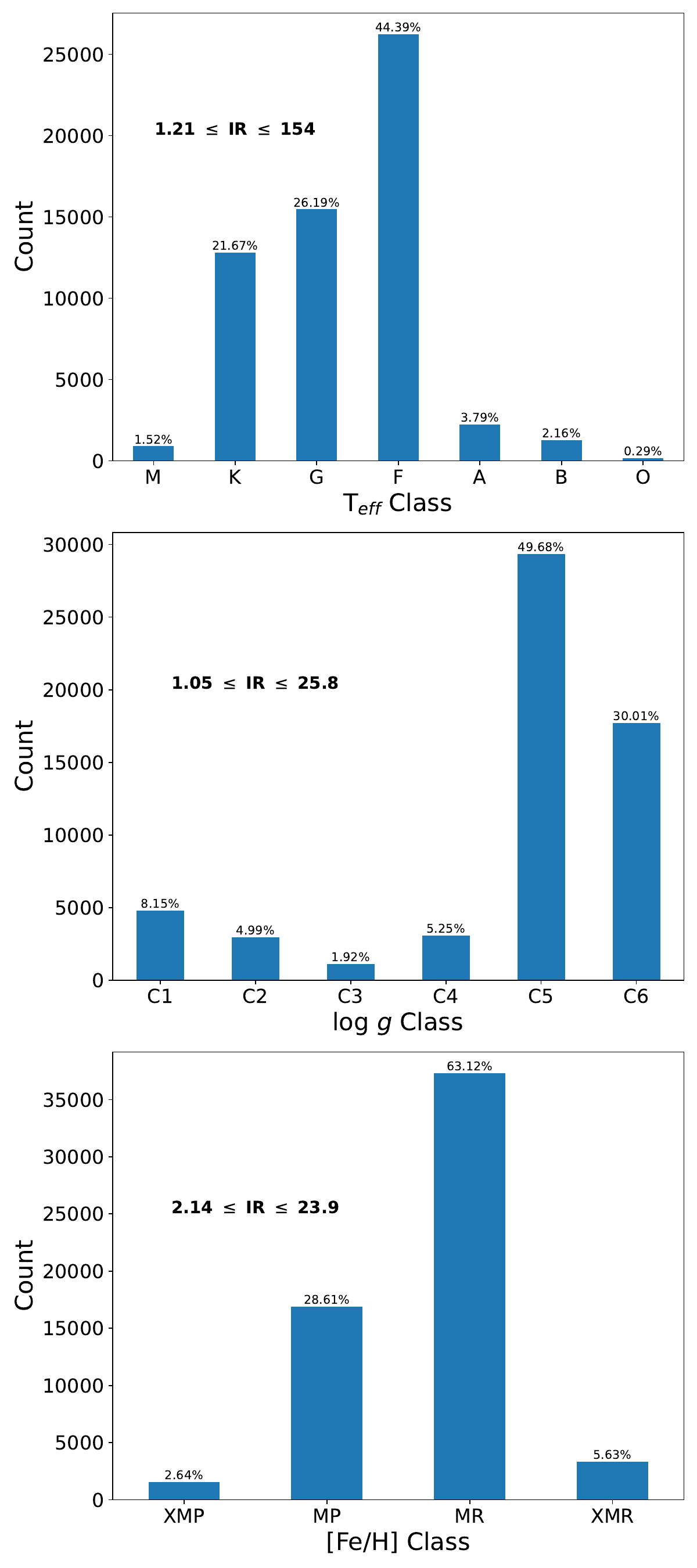}
		\caption{Class distributions for each of the stellar parameters according to the ranges given in Table \ref{tab:class_ranges}, demonstrating high imbalance ratios (IR) for all three parameters.}
		\label{fig:class_distros}
	\end{figure}

\begin{figure}
		\includegraphics[width=\columnwidth]{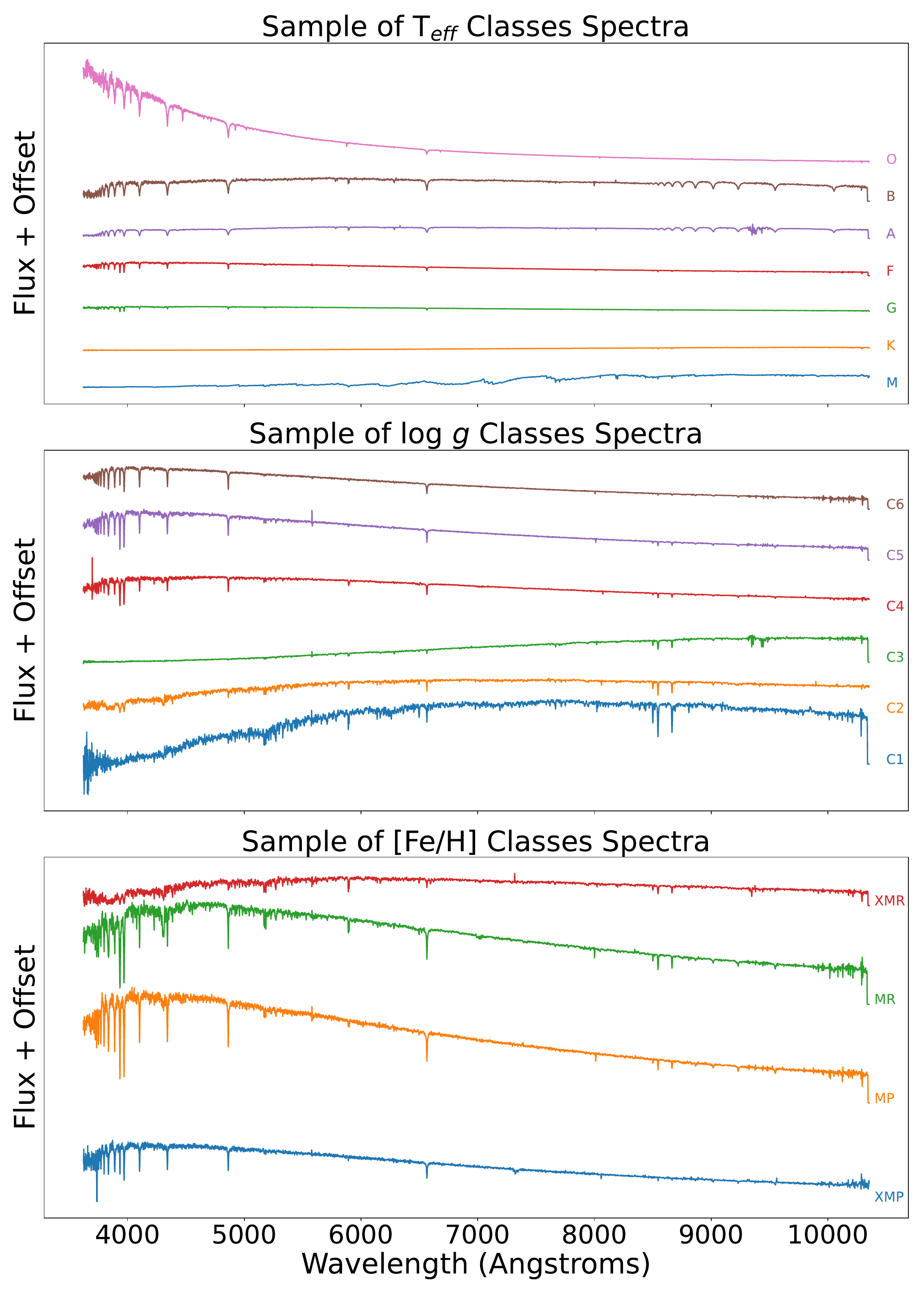}
		\caption{Sample spectra from different classes for each stellar parameter.}
		\label{fig:sample_spectra}
	\end{figure}

	We apply our approach to three stellar parameters: effective temperature ($T_{\text{eff}}$), surface gravity ($\log{g}$), and iron metallicity ([Fe/H]). We only include stars that have a median value available in the VAC for each of these parameters. After dropping unqualified visits, we end up with 59,085 spectra (visits) of 24,162 unique stars. For this set of spectra, more than 85\% have signal-to-noise ratio (S/N) $> 50$, with an overall mean value of about 126. The resulting stellar parameter ranges are as follows:
	\begin{enumerate}
		\item $2,800 \,\text{K} \lessapprox T_{\text{eff}} \lessapprox 31,000 \,\text{K}$,
		\item  $-0.25 \,\text{dex} \lessapprox \log{g} \lessapprox 5.25 \,\text{dex}$,
		\item $-2.75 \,\text{dex} \lessapprox \text{[Fe/H]} \lessapprox 1.00 \,\text{dex}$.
	\end{enumerate}
	The final parameter distribution is also shown in Fig. \ref{fig:mastar_params}. Each of the three parameters was then used separately to classify the spectra into categories according to the ranges shown in \mbox{Table \ref{tab:class_ranges}}. The resulting class distribution for each parameter is shown in Fig. \ref{fig:class_distros}. It is clear that the dataset is highly imbalanced for all three parameters where the imbalance ratio (IR) ranges are as follows:
	\begin{enumerate}
		\item $T_\text{eff}$: 1.21 -- 154,
		\item $\log{g}$: 1.05 -- 25.8,
		\item $\text{[Fe/H]}$: 2.14 -- 23.9.
	\end{enumerate}
	Finally, Fig. \ref{fig:sample_spectra} shows a sample spectrum for every class of each parameter.
	
	\begin{table}
		\caption{Classification ranges for separate stellar parameters, where XMP, MP, MR, and XMR correspond to extremely metal poor, metal poor, metal rich, and extremely metal rich, respectively.}
		\label{tab:class_ranges}
		\centering
		\footnotesize
		\begin{tabular}{lclclc}
			\hline\hline
			\multicolumn{2}{c}{$T_{\text{eff}}$ (K)} & \multicolumn{2}{c}{$\log{g}$ (dex)} & \multicolumn{2}{c}{[Fe/H] (dex)}\\
			Class & Range & Class & Range & Class & Range\\
			\hline
			M & $< 3,500$ & C1 & $< 2.0$ & XMP & $< -2$\\
			K & $3,500\,$--$\,5,000$ & C2 & $2.0\,$--$\,3.0$ & MP & $(-2)\,$--$\,(-1)$\\
			G & $5,000\,$--$\,6,000$ & C3 & $3.0\,$--$\,3.5$ & MR & $(-1)\,$--$\,0$\\
			F & $6,000\,$--$\,7,500$ & C4 & $3.5\,$--$\,4.0$ & XMR & $\ge 0$\\
			A & $7,500\,$--$\,10,000$ & C5 & $4.0\,$--$\,4.5$ &  & \\
			B & $10,000\,$--$\,25,000$ & C6 & $\ge 4.5$ &  & \\
			O & $\ge 25,000$ &  &  &  & \\
			\hline
		\end{tabular}
	\end{table}
	
	\section{Method}
	\label{sec:method}
\begin{figure}
		\centering
		\includegraphics[width=\columnwidth, trim= 20mm 130mm 100mm 25mm, clip]{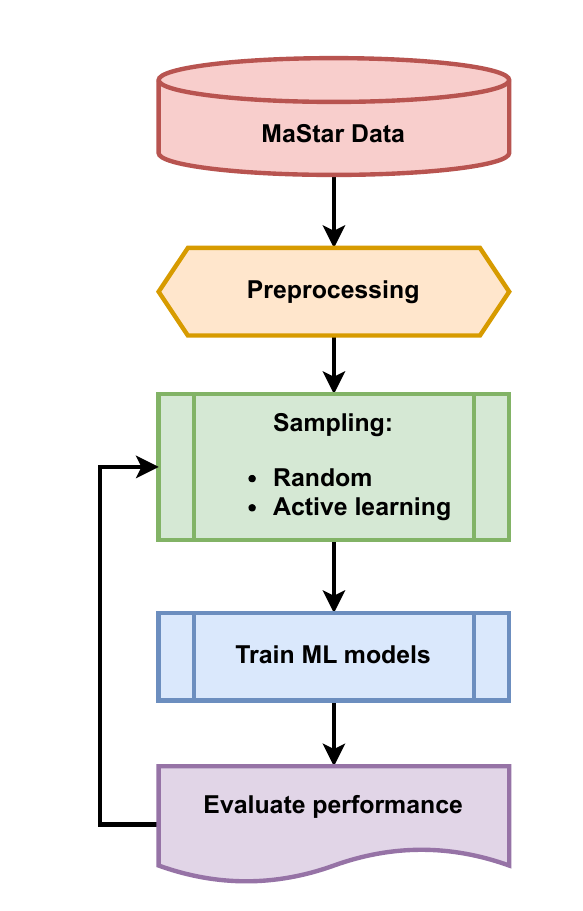}
		\caption{This flowchart outlines the main steps taken in this study.}
		\label{fig:pipeline_flowchart}
	\end{figure}

	In this section, we describe the methods applied in this study. We first prepare the data for use in Section \ref{subsec:preprocessing}. Then, we apply the AL algorithms described in Section \ref{subsec:AL} to curate training samples. The output is iteratively used to train ML models as outlined in Section \ref{subsec:ML}, and the results are compared with a random-sampling benchmark according to the metrics defined in Section \ref{subsec:metrics}. The pipeline of the study is shown as a flowchart in Fig. \ref{fig:pipeline_flowchart}, and further details of the experiments and steps applied are described in Section \ref{subsec:pipeline}. The Python code created for this work is made available on GitHub\footnote{\url{https://github.com/rehamelkholy/StellarAL}}.
	
	\subsection{Preprocessing}
	\label{subsec:preprocessing}
	Before using any dataset to train an ML model, it has to be suitably prepared. To this end, we apply the following four-step preprocessing scheme:
	\begin{enumerate}
		\item we employ a feature-selection routine adapted from the algorithm used by \citet{Brice2019} as a first step toward dimensionality reduction,
		\item split the dataset into training and testing sets,
		\item use min-max normalization to scale each of the selected features, and
		\item apply PCA to further reduce dimensionality.
	\end{enumerate}
	
	Feature selection is a common approach to dimensionality reduction, where we extract the most relevant set of features to reduce the number of dimensions of the input space. It helps with speeding up the algorithm while also discarding some of the noise inherent in the data. There is more than one way to achieve this but we apply the approach proposed by \citet{Brice2019} where we pick flux measurements around specific absorption lines. \citeauthor{Brice2019} included the \ion{H}{\tiny$\updelta$} (4,102 $\,\r{A}$) and \ion{Ca}{i} (4,227 $\r{A}$) lines as they cover six of the seven main spectral classes. The idea is that the flux intensity of such lines is what determines the spectral class while the width of the lines is what determines the luminosity class. Thus, it is not enough to include the flux measurement closest to the wavelength of the absorption line in question; a sufficiently wide region needs to be included around the wavelength to account for the line width in addition to the shifting of the spectrum due to radial velocity. We adopt the same procedure, but since our model is applied not only to spectral and luminosity classes but also to metallicity classes, instead of only using the two lines mentioned above, we include additional lines as listed below:
	\begin{enumerate}
		\item \ion{Ca}{ii k} (3,934 $\r{A}$) and \ion{Ca}{ii h} (3,968 $\r{A}$): Key indicators in A-type stars, showing strength variances linked to temperature and luminosity effects \citep{Gray2009},
		\item \ion{Fe}{i} (4,046 $\r{A}$): Often used with the hydrogen line H\thinspace{\tiny{$\updelta$}} for temperature classification, especially in F-type and later-type stars \citep{Gray2009},
		\item Two spectral lines can help us classify stars from B to M \citep{Brice2019}:
		\begin{itemize}
			\item \ion{H}{\tiny{$\updelta$}} (4,102 $\r{A}$): Present in B-, A-, F-, and G-type stars;
			\item  \ion{Ca}{i} (4,227 $\r{A}$): Present in F-, G-, K-, and M-type stars,
		\end{itemize}
		\item G-band CH (4,300 $\r{A}$): Prominent in late-G to K-type stars and is sensitive to surface gravity \citep{Gray2009},
		\item \ion{He}{ii} (4,686 $\r{A}$): Dominates O-type stars \citep{Gray2009},
		\item TiO band (4,955 $\r{A}$): At least one TiO band is needed for M-type classification \citep{Gray2014},
		\item \ion{Fe}{i} (5,269 $\r{A}$) and \ion{Fe}{ii} (5,018 $\r{A}$): For metallicity classification \citep{Santos2004}.
	\end{enumerate}
	Table \ref{tab:abs_lines} lists the set of spectral lines used, the central wavelength corresponding to each of them, and the wavelength range included to account for the spectral line in the reduced feature space. The flux measurements from all regions are combined at the end to create one flux array per spectrum. At this preprocessing step, we reduce the feature-space dimensionality from 4,563 to 170.
	
	\begin{table}
		\caption{The absorption lines included in the feature selection step, their central wavelengths, and feature ranges in ascending order, with a consistent wavelength spacing of about 0.83 $\r{A}$.}
		\label{tab:abs_lines}
		\centering
		\footnotesize
		\begin{tabular}{lcc}
			\hline\hline
			Line 						& Central Wavelength ($\r{A}$) 	& Wavelength Range ($\r{A}$)\\
			\hline
			\ion{Ca}{ii k}				& 3,934							& 3,926.45 -- 3,940.94\\
			\ion{Ca}{ii h}				& 3,968							& 3,960.96 -- 3,975.58\\
			\ion{Fe}{i}					& 4,046							& 4,038.31 -- 4,053.22\\
			\ion{H}{\tiny$\updelta$}	& 4,102							& 4,094.49 -- 4,109.60\\
			\ion{Ca}{i}					& 4,227							& 4,218.91 -- 4,234.48\\
			G-band CH					& 4,300							& 4,292.40 -- 4,308.24\\
			\ion{He}{ii}				& 4,686							& 4,677.35 -- 4,694.62\\
			TiO band					& 4,955							& 4,945.38 -- 4,963.64\\
			\ion{Fe}{ii}				& 5,018							& 5,008.41 -- 5,026.90\\
			\ion{Fe}{i}					& 5,269							& 5,258.96 -- 5,278.37\\
			\hline
		\end{tabular}
	\end{table}
	
	Since the last two preprocessing steps of the scheme outlined above include parameter fitting, the data has to be split first as only training data can be used in the fitting process in order to prevent data leakage. Accordingly, 10\% of the dataset is set aside for testing. Because the dataset is highly imbalanced, stratification during data splitting is necessary to ensure that the evaluation metrics obtained at the testing step accurately reflect the model performance. Moreover, to avoid any ambiguity that might result from multi-label stratification, this step is applied separately to a copy of the entire dataset for each of the three classification parameters: $T_{\text{eff}}$, $\log{g}$, and [Fe/H]. This leaves us with 53,176 samples for training and 5,909 for testing.
	
	After splitting the dataset into training and testing sets, each feature is scaled using the equation:
	\begin{equation}
		f_{i,\text{scaled}} = \dfrac{f_i - f_{i,\text{min}}}{f_{i,\text{max}} - f_{i,\text{min}}},
	\end{equation}
	where $f_i$ is the $i^{\text{th}}$ flux measurement, $f_{i,\text{max}}$ and $f_{i,\text{min}}$ are the corresponding maximum and minimum flux measurements, respectively, and $f_{i,\text{scaled}}$ is the scaled flux measurement. This step is crucial to provide a frame-of-reference for the model to compare feature values for different samples. However, as mentioned before, only the training set can be used in determining the minimum and maximum values for each feature. These values are then used to scale the testing set. This process is easily handled by use of the \texttt{MinMaxScaler} of the \texttt{sklearn.preprocessing} module \citep{Pedregosa2011}, where the scaler is first fit by the training set and later used to transform the entire dataset.
	
\begin{figure}
		\centering
		\includegraphics[width=\columnwidth]{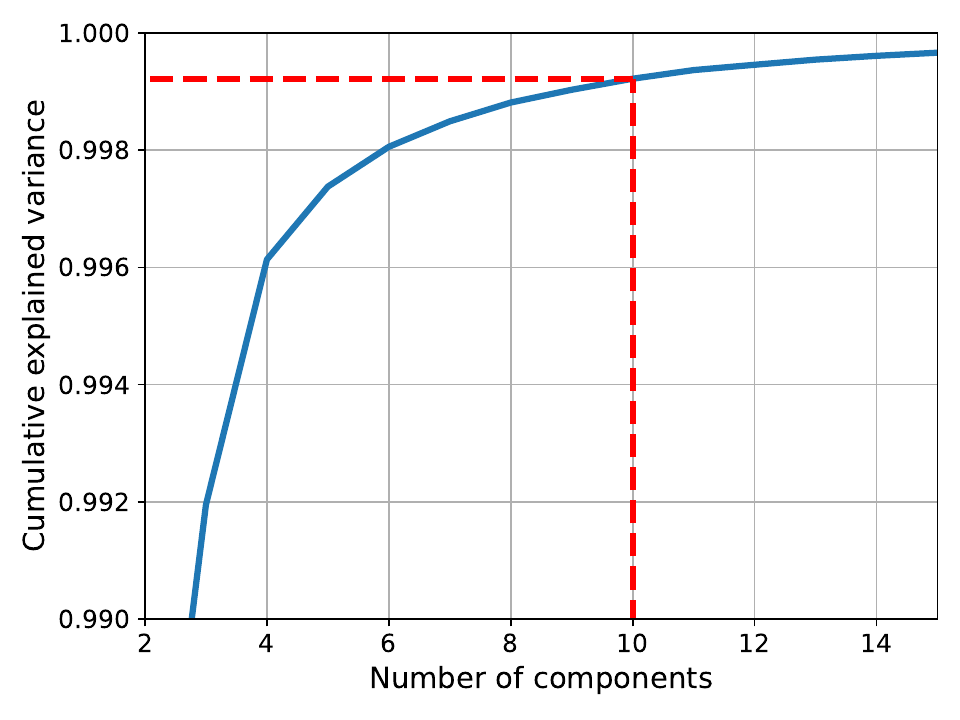}
		\caption{The cumulative explained variance plotted against the number of PCA components, showing that more than 99.95\% of the variance is explained by the first 9 components.}
		\label{fig:PCA}
	\end{figure}

	Due to the higher computational expense of the AL algorithms described in Section \ref{subsec:AL}, we had to minimize the number of features further. Thus, we use PCA, which is a statistical method that defines a linear transform to reduce a dataset to its most essential features (i.e. principal components). These components are ordered according to the variance captured by each of them. Using this transform, an approximation of the dataset can be obtained by a few major components. A thorough description of the PCA method can be found in \citet{Ivezic2020} or \citet{Greenacre2022}. In this work, we first apply PCA to the entire dataset (after applying the feature scaling routine described above) to determine the number of principal components we will include in our model. As shown in Fig. \ref{fig:PCA}, we found that more than 99.95\% of the variance in the data is captured by the first 9 components. Early trials also indicated that a higher number of features results in an increase in computational cost that cannot be justified by any improvement in the model performance. After the dataset has been split and scaled separately for each parameter, PCA is applied on the training sets and the resulting approximations are used to map the testing sets as well. This process was practically executed using the \texttt{PCA} class from the \texttt{sklearn.decomposition} module \citep{Pedregosa2011}.
	
	\subsection{Active learning approach}
	\label{subsec:AL}
	The role of a classification ML model is essentially to generate a mapping between input features and the class labels based on the features and labels of the training dataset. However, for this mapping to be as accurate as possible, large amounts of labelled training instances are required. The labelling process is often very expensive in terms of time and manpower. The collection of such data is currently one of the main challenges in ML applications \citep{Dimitrakakis,Li2021}. The solution to this problem would be to minimize the size of the needed training data while only keeping the most high-quality data. This could be achieved by careful selection of unlabelled instances to later be labelled by an annotator or expert, which is the goal of \emph{Active Learning} (AL) \citep{Huang2014,Ghahramani2017}.
	
\begin{figure}
		\centering
		\includegraphics[width=\columnwidth, trim= 9mm 6mm 9mm 4mm, clip]{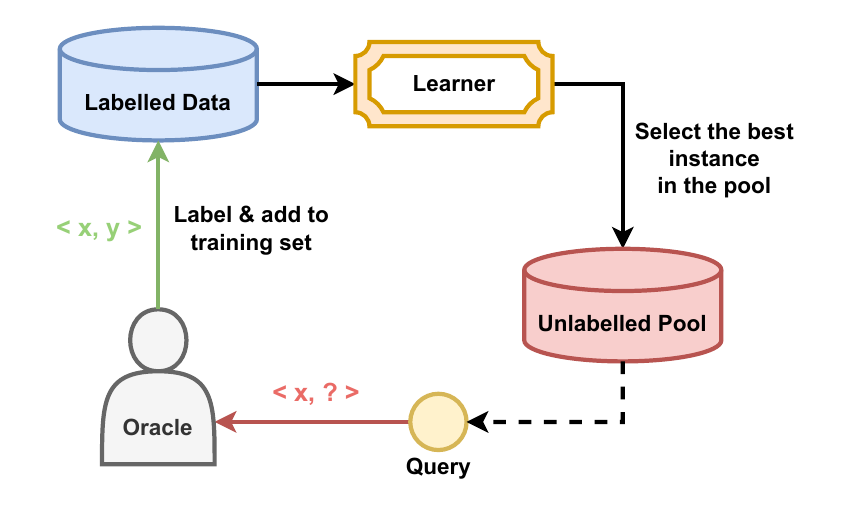}
		\caption{An illustration of the pool-based active-learning scenario querying the most informative instance from a large pool of unlabelled data.}
		\label{fig:pool_al}
	\end{figure}

	AL algorithms can be categorized into three main scenarios: membership query synthesis, stream-based selective sampling, and pool-based active learning; which is the most well-known of them and to which the algorithms we use in this work belong. A detailed discussion of the three different scenarios and the advantages and limitations of each can be found in \citet{Settles2012} or \citet{Tharwat2023}. The pool-based sampling approach selects instances from an existing pool of unlabelled data based on the active learner evaluation of the informativeness of some or all of the instances in the pool. The selected instance is then annotated by the oracle and added to the labelled training set. This process is iteratively repeated until a criterion is reached, which is usually a maximum number of iterations. Thus, this type of scenario generally includes two adjustable parameters: the initial labelled batch size and the number of additional instances to be queried. Fig. \ref{fig:pool_al} illustrates the pool-based sampling approach. In this work, we tested six different sampling strategies that can be divided into two categories. The first category is \emph{uncertainty sampling} which includes three strategies based on three uncertainty measures: classification uncertainty, classification margin, and classification entropy. The second category is \emph{query by committee} (QBC) which includes three strategies as well based on three disagreement measures: vote entropy, consensus entropy, and maximum disagreement. We give a brief definition of each strategy below, but a more thorough explanation can be found in \citet{Settles2012}.
	
	On one hand, uncertainty sampling evaluates each instance in the unlabelled pool and presents the most informative one to be annotated and added to the labelled training set, where the evaluation of instances is based on an uncertainty measure, hence the name. The first measure we try here is classification uncertainty defined by:
	\begin{equation}
		U(x) = 1 - P(\hat{x}|x),
	\end{equation}
	where $x$ is the instance to be predicted and $\hat{x}$ is the most likely prediction. The strategy selects the instance with the highest uncertainty. For the classification margin strategy, the difference in probability between the first and second most likely classes is calculated according to:
	\begin{equation}
		M(x) = P(\hat{x}_1|x) - P(\hat{x}_2|x),
	\end{equation}
	where $\hat{x}_1$ and $\hat{x}_2$ are the first and second most likely classes, respectively. In this case, the strategy selects the instance with the smallest margin, since it means that the learner is less decisive about the predicted class. Finally, classification entropy is calculated using:
	\begin{equation}
		H(x) = - \underset{k}{\sum} p_k \log(p_k),
	\end{equation}
	where $p_k$ is the probability of the sample belonging to the $k^\text{th}$ class. This is proportional to the average number of guesses that has to be made to find the true class. Thus, the strategy selects the instance with the largest entropy.
	
	On the other hand, QBC strategies are based on having several hypotheses (i.e. classifiers) about the data, and querying the instances based on measures of disagreement between the hypotheses. The first measure we try is vote entropy defined by:
	\begin{equation}
		E_{\text{vote}}(x) = - \underset{y}{\sum} \dfrac{N(y|x)}{|\mathcal{C}|} \log{\dfrac{N(y|x)}{|\mathcal{C}|}},
	\end{equation}
	where $N(y|x)$ is the number of `votes' the class $y$ receives for instance $x$ among the hypotheses in committee $\mathcal{C}$, and $|\mathcal{C}|$ is the committee size. This strategy selects the instance where $E_{\text{vote}}$ is the largest, since it corresponds to the most uniform distribution of votes among classes. It is a `hard' vote entropy measure; we also try a `soft' vote entropy measure referred to as consensus entropy which accounts for the confidence of each committee member and is defined by:
	\begin{equation}
		E_{\text{cons}}(x) = - \underset{y}{\sum} P(y|x) \log{P(y|x)},
	\end{equation}
	where $P(y|x)$ is the average `consensus' probability that $y$ is the correct class according to the committee. Finally, the maximum disagreement measure is based on the \emph{Kullback-Leibler} (KL) \emph{divergence} \citep{Kullback1951}, which is a measure of the difference between two probability distributions. In other words, the disagreement is quantified as the average divergence of each classifier's prediction from that of the consensus $\mathcal{C}$ as follows:
	\begin{equation}
		D(x) = \dfrac{1}{|\mathcal{C}|} \underset{\theta \in \mathcal{C}}{\sum} \text{KL} \left( P_{\theta}(Y|x) \parallel P_{\mathcal{C}}(Y|x) \right),
	\end{equation}
	where the KL divergence of committee member $\theta$ is defined by:
	\begin{equation}
		\text{KL} \left( P_{\theta}(Y|x) \parallel P_{\mathcal{C}}(Y|x) \right) = \underset{y}{\sum} P_\theta (y|x) \log{\dfrac{P_\theta (y|x)}{P_{\mathcal{C}} (y|x)}}.
	\end{equation}
	As the name suggests, this strategy picks the instance with the maximum disagreement value, $D_\text{max}$.
	
	In this work, we use each of the AL strategies described above to iteratively sample instances for training supervised ML models. We apply this approach to each of the three stellar parameters separately, taking random sampling as a baseline for comparison. We use the Modular Active Learning framework for Python3 (\texttt{modAL}) \citep{Danka2018} to implement these strategies directly into our code. The pipeline of the entire experimental steps is detailed in Section \ref{subsec:pipeline}.
	
	\subsection{Machine learning models}
	\label{subsec:ML}
	In this work, we use different supervised-learning algorithms and compare their performances according to the metrics described in Section \ref{subsec:metrics}. We apply three ML models: $k$-nearest neighbours (KNN), random forest (RF) \citep{Breiman2001}, and gradient boosting (GB) \citep{Friedman2001}; in addition to an ensemble model that combines their outputs. Some ML algorithms are only used in certain experiments as detailed in Section \ref{subsec:pipeline}. In what follows, we briefly introduce each algorithm.
	
	\paragraph*{KNN}~\\
	The $k$-nearest neighbours is a clustering algorithm based on distance metrics. The standard Euclidean distance is most-commonly chosen as the distance metric measure. KNN can be applied to both regression and classification problems. Since it is one of the simpler algorithms, it offers a robust way to establish a baseline for classification accuracy. At its core, it is based on the assumption that if two data points are nearby each other, they belong to the same class. $k$ is a tunable parameter that represents the number of neighbouring points to be considered, such that the classification of a data point relies on the voting results of the $k$ neighbours that are nearest to it in the multidimensional space.
	
	\paragraph*{Random Forest}~\\
	RF \citep{Breiman2001} is widely used for classification and regression problems because it is fast to train and scales well while also maintaining competitive performance to other ML algorithms. RF is an ensemble method consisting of randomly-generated decision trees. It uses bootstrap sampling techniques which means that different decision trees are simultaneously trained on different subsets of the training data using random subsets of the features. Thus, while a decision tree usually overfits, RF is less prone to overfitting as it uses the average of the trees, which ultimately improves classification accuracy.
	
	\paragraph*{Gradient Boosting}~\\
	The GB algorithm \citep{Friedman2001} is a powerful ensemble model that combines multiple decision trees to create a stronger predictive model. In GB, each subsequent tree corrects the errors of the previous one. It optimizes a specific objective function, typically a loss function, by minimizing it through gradient descent. Overall, GB achieves higher performance and better generalization with lower time cost than other ensemble learning methods, such as the stochastic forest algorithm and the support vector machine (SVM) of a single model \citep{Zeraatgari2023}.
	
	\paragraph*{Voting}~\\
	Voting is an ensemble learning technique used in classification and regression tasks. It combines predictions from multiple base classifiers and selects the class label by voting, which leads to improved performance compared to individual classifiers. The voting classifier can be implemented using soft or hard voting. A hard voting classifier chooses the class with the highest frequency of votes, whereas a soft voting one averages the class probabilities across all base classifiers. Different base classifiers can be given different voting weights based on their individual performances. In this work, we combine KNN, RF, and GB in a soft voting classifier, giving RF and GB weights of 2 each and KNN a weight of 1.
	
	\subsection{Metrics}
	\label{subsec:metrics}
\begin{figure}
		\centering
		\includegraphics[width=\columnwidth]{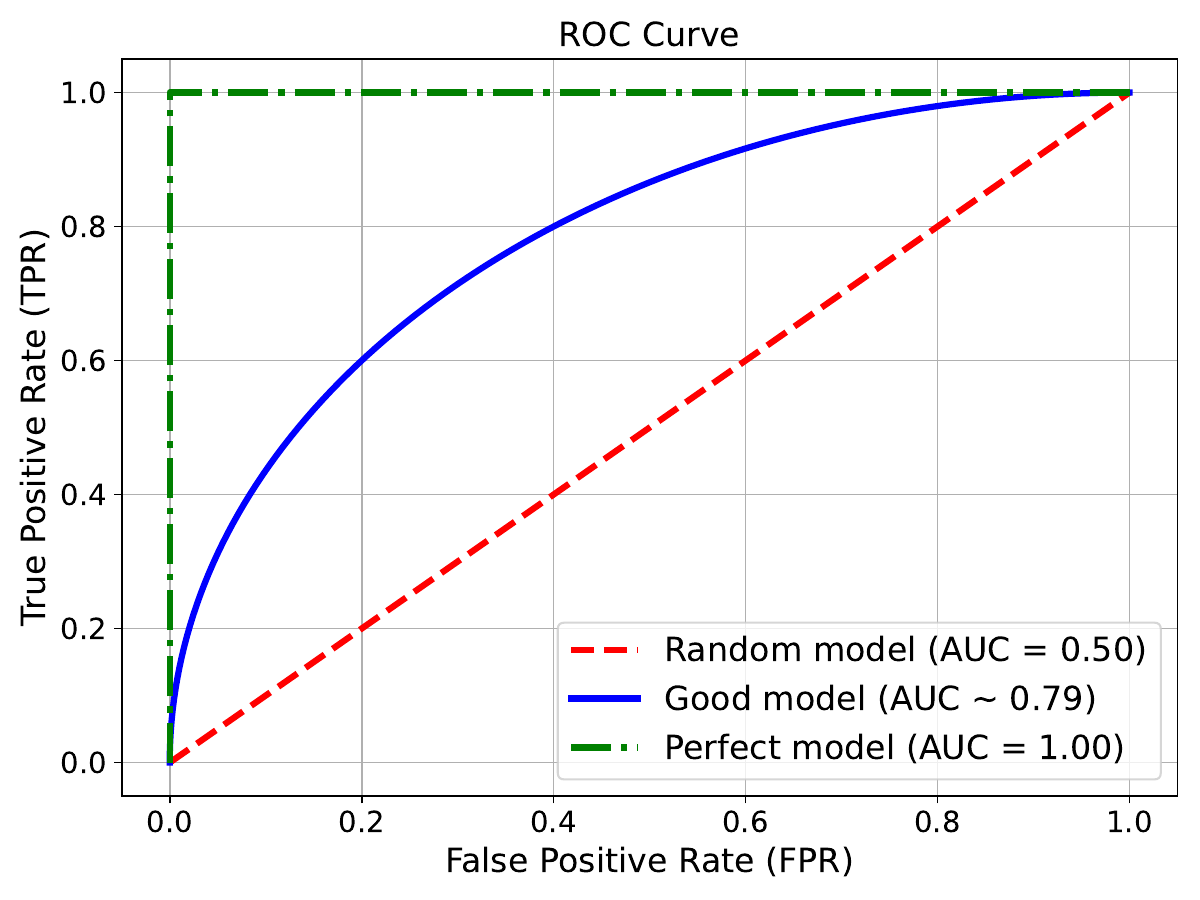}
		\caption{Examples of different Receiver Operating Characteristic (ROC) for models with different levels of performances, where the legend shows the area under the curve (AUC) value for each model.}
		\label{fig:ROC}
	\end{figure}

	In this study, we aim to compare the performances of different sampling methods, in addition to the performances of ML models. Accuracy is the most basic evaluation metric, and is given by:
	\begin{equation}
		\text{Accuracy} = \dfrac{\text{TP} + \text{TN}}{\text{TP} + \text{FP} + \text{TN} + \text{FN}},
	\end{equation}
	where TP, FP, TN, and FN are the numbers of true positives, false positives, true negatives, and false negatives, respectively. However, accuracy does not give an accurate reflection of the model performance in the case of class imbalance. Hence, a more helpful pair of metrics can be employed for that; namely, sensitivity and specificity. Sensitivity, or the true positive rate (TPR), measures the ability of a model to classify positives correctly, and is given by:
	\begin{equation}
		\text{Sensitivity} = \dfrac{\text{TP}}{\text{TP} + \text{FN}},
	\end{equation}
	while specificity, or the true negative rate (TNR), measures the ability of a model to classify negatives correctly. and is given by:
	\begin{equation}
		\text{Specificity} = \dfrac{\text{TN}}{\text{TN} + \text{FP}}.
	\end{equation}
	It is also conventional to use another metric that relates both sensitivity and specificity, which is the area under the curve (AUC) of the Receiver Operating Characteristic (ROC) curve, where TPR is plotted against the false positive rate (FPR) at different threshold values, where FPR is given by:
	\begin{equation}
		\text{FPR} = \dfrac{\text{FP}}{\text{FP} + \text{TN}} = 1 - \text{specificity}.
	\end{equation}
	AUC is a measure of the total two-dimensional area under the ROC curve; and is a very good predictor of the overall performance of the classifier, where a baseline random model is expected to have an AUC $\sim 0.5$ and a perfect model would have an AUC value of 1. Fig. \ref{fig:ROC} shows examples of ROC curves for models with different levels of performances.
	
	For a multi-class imbalanced dataset, sensitivity is of particular interest since it emphasises the ability of the model to correctly identify true positives of minority classes; whereas a model can score high specificity even if it can only classify the majority classes. To take that into account, we use all three metrics to compare the AL algorithms with random sampling. For all three metrics, we calculate the \emph{macro} value for the metrics. That is, we evaluate the metric for each class separately and take the average as the final metric value. This is an added measure to give the performance of the model on minority classes the same weight as its performance on majority ones.
	
	In addition, we use the Matthew’s Correlation Coefficient (MCC) as a fourth metric. It is defined by:
	\begin{equation}
		\text{MCC} = \dfrac{\text{TP} \times \text{TN} - \text{FP} \times \text{FN}}{\sqrt{(\text{TP} + \text{FP}) (\text{TP} + \text{FN}) (\text{TN} + \text{FP}) (\text{TN} + \text{FN})}}.
	\end{equation}
	Because MCC uses all four elements of the confusion matrix (TP, TN, FP, and FN) in the numerator, it does not get skewed by class imbalance, which makes it a more reliable metric for summarizing the overall performance of the model across all classes. MCC ranges from $-1$ (total disagreement between predicted and true labels) to $1$ (perfect prediction), where $\text{MCC} = 0$ indicates a near-random prediction. This makes it a very intuitive metric for understanding the performance of a classifier. A more thorough explanation of each of the chosen metrics can be found in \citet{Marsland2014}.
	
	Even though we defined the metrics above in terms of binary classification to provide a clear concept of what they measure, these definitions are easily generalized to fit their application on a multi-class dataset. This is easily handled by making use of the \texttt{imblearn.metrics} module \citep{Lemaitre2017} to calculate both sensitivity and specificity, and the \texttt{sklearn.metrics} module \citep{Pedregosa2011} for the AUC and MCC.
	
	\subsection{Pipeline}
	\label{subsec:pipeline}
	In what follows we outline the steps taken in the experiments performed in this work in order. The aim of the first experiment carried out was to compare the performances of models trained on samples queried by different approaches. Since the aim is not to optimize the ML model selected but the sampling algorithm, we started by training each of the models described in Section \ref{subsec:ML} on the entire dataset to select the highest-performing one. The chosen model was then used throughout the rest of the steps, except for QBC strategies, where we use a committee of three learners; namely, KNN, RF, and GB, all initialized using the same batch. We then evaluated each sampling strategy, separating uncertainty sampling strategies from QBC strategies and using random sampling as a baseline in both cases. We varied the initial batch size, taking care to initialize all strategies using the same batch. We begin by evaluating the initial model on the testing set before iteratively augmenting the training sample by querying the data pool and retraining the model. The model performance is reevaluated after each 5 queries using each of the metrics listed in Section \ref{subsec:metrics}. We perform 20 runs as described to account for the performance variance for some strategies, and calculate the mean performance metrics for all runs. This experiment is repeated for each of the stellar parameters separately.
	
	As a final step to statistically assess the impact of AL strategies on performance, we aggregate the results across all runs and batch sizes for uncertainty-sampling and QBC strategies separately and apply the \emph{Wilcoxon signed-rank test} \citep{Wilcoxon1992} on each metric, repeating the steps for each stellar parameter. We set a predefined significance threshold of $p<0.05$. In addition, we calculated the associated values of the \emph{Cohen's $d$} impact index \citep{Cohen2013}, which is often used to quantify the practical significance of the difference between two means. Values of 0.2 to 0.5 indicate low significance, 0.5 to 0.8 indicate medium significance, and greater than 0.8 indicate large significance.
	
	The aim of the second experiment is to assess the performance progress of a model trained on an AL-sampled set with the increase of the number of additional instances. This experiment is only carried out on effective temperature. We pick the highest-performing strategy from the first experiment to use with the same best-performing model chosen before, and only sensitivity is used to assess the models in this experiment. For the sake of comparison, we use three baseline training sets:
	\begin{enumerate}
		\item the whole initial training set,
		\item a random sample of 10\% of the initial training set,
		\item a stratified sample of 10\% of the initial training set.
	\end{enumerate}
	For both (2) and (3), 20 different samples are used to account for performance variance and the mean results are calculated along with their standard deviations. Finally, we run the AL-sampling method 5 times (averaged at the end) to sample 5\% of the initial training data pool and retrain and reevaluate the model every 5 queries. 
	
	\section{Results and discussion}
	\label{sec:results}
	In this section, we present and discuss the results of the experiments carried out in this study. We began by evaluating the performance of different ML models on the testing set after being trained on the entire training set. The results of these steps are shown in \mbox{Table \ref{tab:ML_perf}}. It can be seen that RF outperforms the other three across all metrics, particularly sensitivity, for both effective temperature and surface gravity. For iron metallicity, the voting models has a slightly better AUC score compared to RF, but the latter still scores higher on the other three metrics. It is also worth noting that the computational cost of the voting model is almost three times that of RF, since it is training the three member learners under the hood. Hence, we decided to use RF for all three parameters moving forward, except when using QBC strategies as mentioned before.
	
	\begin{table}
		\caption{Performance scores for different ML models when evaluated on the testing set after being trained on the entire training set for the three stellar parameters.}
		\label{tab:ML_perf}
		\centering
		\footnotesize
		\begin{tabular}{lccccc}
			\hline\hline
			Parameter			& Model		& AUC   & MCC	& Sensitivity	& Specificity\\
			\hline
			$T_{\text{eff}}$	& KNN    	& 0.947	& 0.876	& 0.787			& 0.982\\
			& RF     	& 0.989	& 0.907	& 0.859			& 0.987\\
			& GB     	& 0.955	& 0.836	& 0.766			& 0.977\\
			& Voting 	& 0.989	& 0.891	& 0.825			& 0.985\\
			&		  	&		&		&				&	   \\
			$\log{g}$		 	& KNN	  	& 0.884	& 0.591	& 0.629			& 0.929\\
			& RF	  	& 0.949	& 0.683	& 0.707			& 0.942\\
			& GB     	& 0.928	& 0.627	& 0.647			& 0.932\\
			& Voting 	& 0.948	& 0.673	& 0.683			& 0.939\\
			&		  	&		&		&				&	   \\
			Fe/H			 	& KNN	  	& 0.920	& 0.772	& 0.711			& 0.939\\
			& RF	  	& 0.979	& 0.856	& 0.787			& 0.961\\
			& GB     	& 0.975	& 0.836	& 0.776			& 0.956\\
			& Voting 	& 0.980	& 0.845	& 0.781			& 0.958\\
			\hline
		\end{tabular}
	\end{table}
	
	We can also see from Table \ref{tab:ML_perf} that the KNN model has the lowest overall scores across all three stellar parameters. This is because we do not perform any hyperparameter tuning for the ML models used, but rather keep the default values of the \texttt{Scikit-learn} library \citep{Pedregosa2011}. In the case of KNN, the most important hyperparameter is the number of neighbours, $k$, which has a default value of 5. Early trials with hyperparameter grid searches for some of the ML models used in this study indicated that a value of 18 achieves better performance scores for the KNN model. This suggests that the overpopulation of the feature space with majority classes makes it necessary to increase the number of neighbours taken into account to correctly classify instances that belong to minority classes.
	
\begin{figure*}
		\centering
		\includegraphics[width=\textwidth]{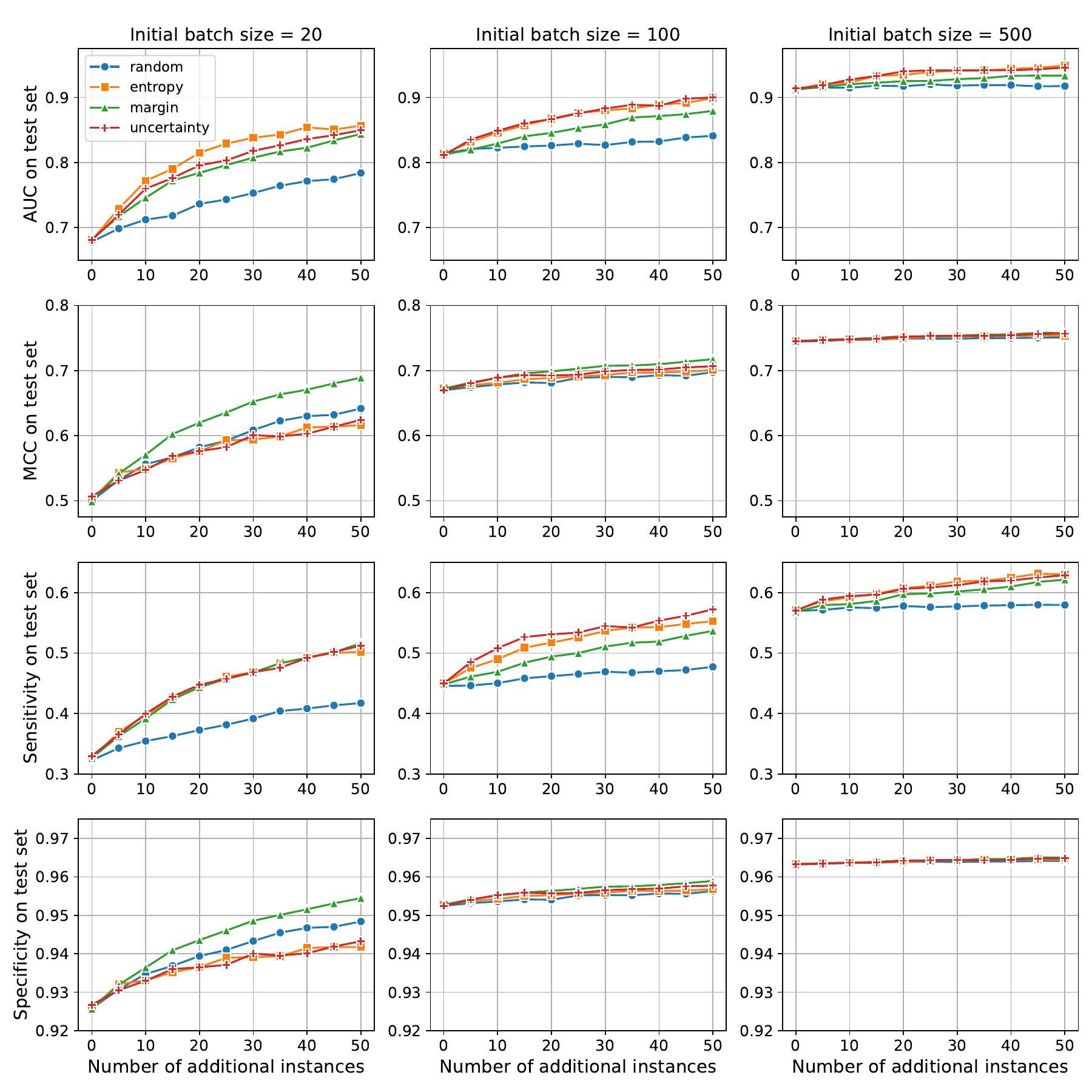}
		\caption{Performance scores for uncertainty sampling strategies along with random sampling for different initial batch sizes applied to $T_{\text{eff}}$.}
		\label{fig:teff_non_comm}
	\end{figure*}

	Fig. \ref{fig:teff_non_comm} shows the performance scores of different single-learner AL sampling strategies along with a random-sampling baseline for different initial batch sizes when applied to effective temperature. It can be seen that, for all metrics, at least one uncertainty sampling strategy outperforms random sampling. In particular, all AL strategies significantly outperform the random baseline on sensitivity scores across all initial batch sizes. The best-performing strategy is clearly the classification margin strategy, but only for an initial batch size ($n_{\text{init}}$) $= 20$. When the first and second columns of subplots in the figure are compared, we can notice that the performance scores after adding 50 AL-sampled instances to an initial randomly-chosen set of 20 is always higher than the corresponding scores when using an initial set of 100 randomly-chosen instances. This demonstrates the effectiveness of AL sampling in achieving better scores with fewer training instances. We can also see that the larger the size of the initial training batch, the less pronounced the improvement in performance due to AL sampling. However, the improvement in sensitivity for all AL strategies is still evident, even with an initial batch of 500 instances. This offers a contrast with the plateauing of random-sampling sensitivity at the same initial batch size. Of course, the emphasis on sensitivity scores is due to the highly-imbalanced nature of the dataset, which makes sensitivity scores more representative of a model's ability to correctly identify minority classes.
	
\begin{figure*}
		\centering
		\includegraphics[width=\textwidth]{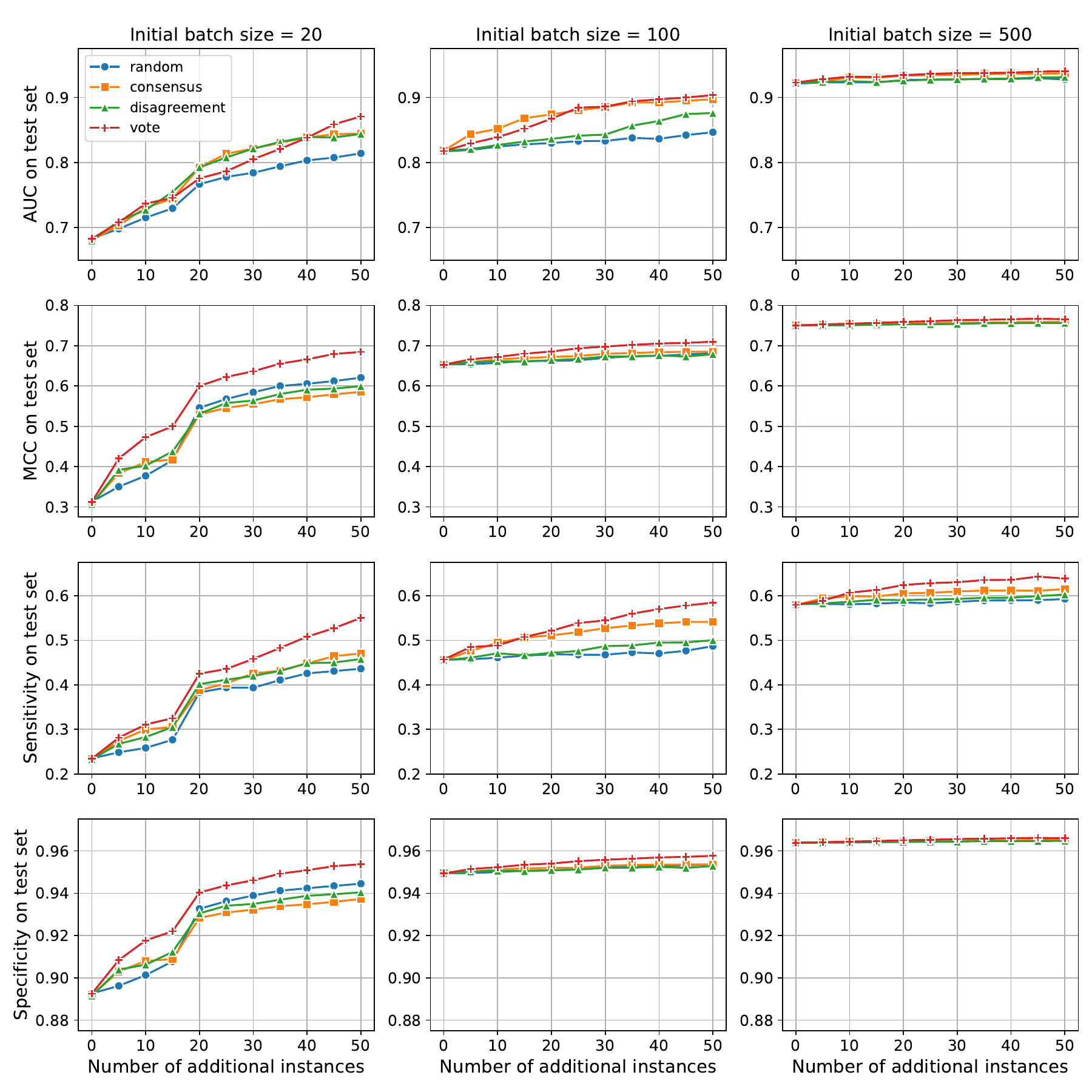}
		\caption{Performance scores for QBC disagreement sampling strategies along with random sampling for different initial batch sizes applied to $T_{\text{eff}}$.}
		\label{fig:teff_comm}
	\end{figure*}

	Fig. \ref{fig:teff_comm} shows the performance scores of different QBC disagreement sampling strategies, along with a random sampling baseline adapted for QBC learning as well, for different initial batch sizes applied to $T_{\text{eff}}$. We can see that some of the scores in this case are higher than those of the single RF model shown in Fig. \ref{fig:teff_non_comm}. However, when we take into account the fact that the computational cost of a QBC model is almost linearly dependent on the number of learners in the committee (three in this case), the corresponding improvement in performance diminishes. Comparing the scores of QBC disagreement strategies with the random approach yields similar results to non-committee uncertainty sampling comparison with random sampling. Nevertheless, it is worth reiterating that AL strategies score higher than random sampling on sensitivity, even after increasing the initial batch size. It is clear that the vote-entropy strategy outperforms all others across all metrics. If computational resources were no issue, higher scores can be obtained via pre-calibration of single committee members. However, this again raises the need for labelled instances to perform such calibration prior to training.
	
\begin{figure*}
		\centering
		\includegraphics[width=\textwidth]{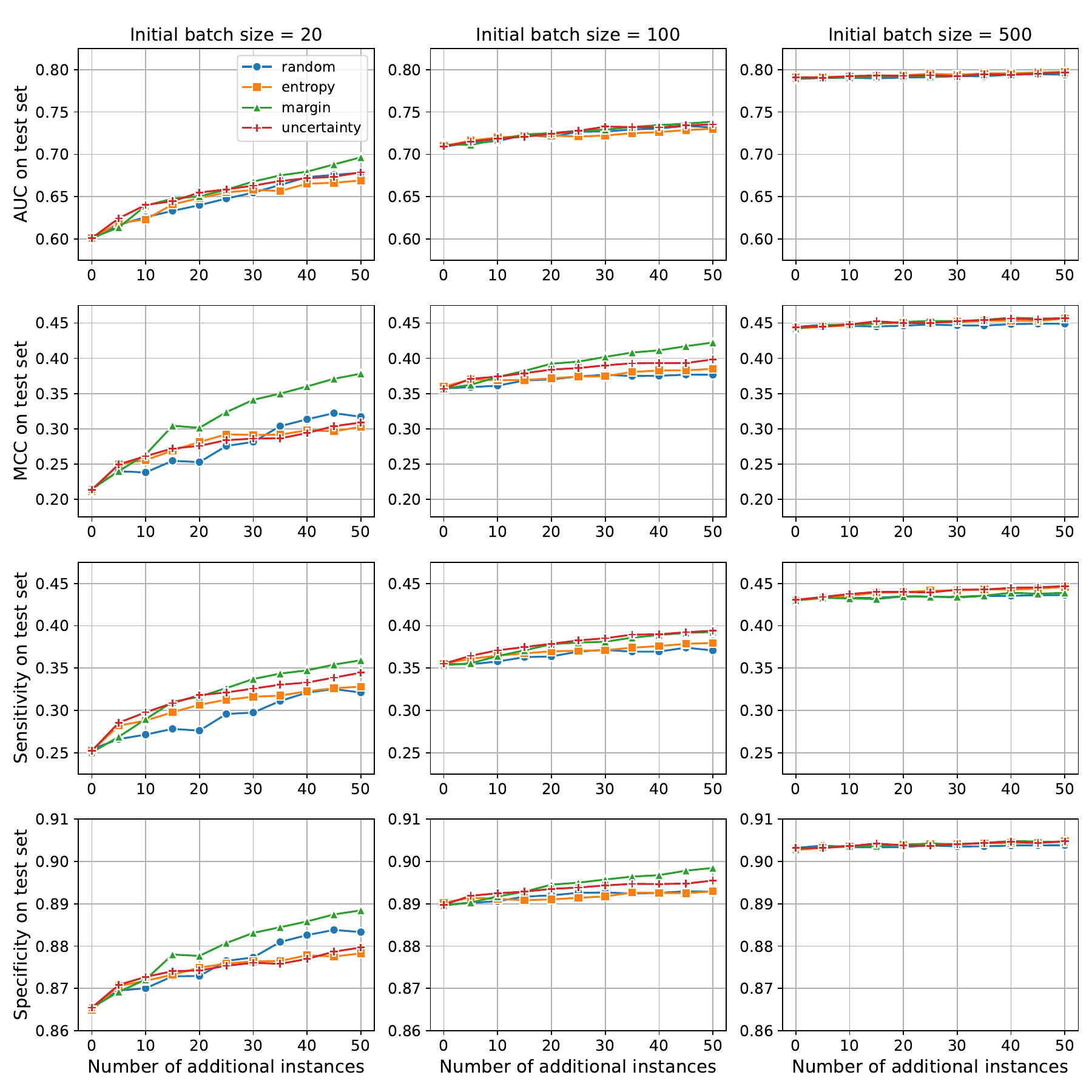}
		\caption{Performance scores for uncertainty sampling strategies along with random sampling for different initial batch sizes applied to $\log{g}$.}
		\label{fig:logg_non_comm}
	\end{figure*}

\begin{figure*}
		\centering
		\includegraphics[width=\textwidth]{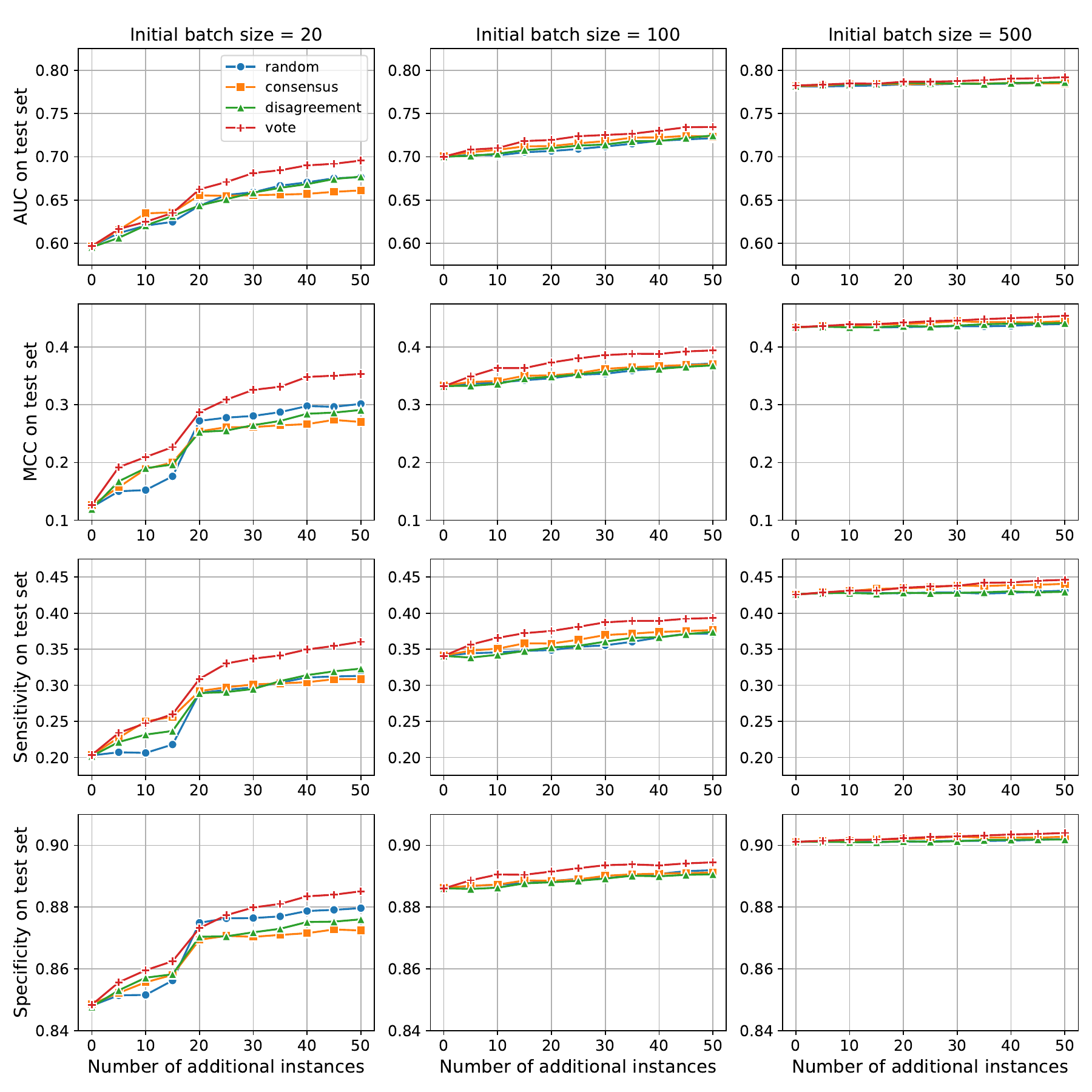}
		\caption{Performance scores for QBC disagreement sampling strategies along with random sampling for different initial batch sizes applied to $\log{g}$.}
		\label{fig:logg_comm}
	\end{figure*}

	The performance scores of uncertainty sampling strategies compared with random sampling applied to surface gravity with different initial batch sizes are shown in Fig. \ref{fig:logg_non_comm}. Random sampling performance is comparable to AL uncertainty sampling strategies for most metrics. However, the classification margin strategy outperforms all the others across all metrics, even with an increasing initial batch size. The differences in MCC and sensitivity scores between margin sampling and random sampling particularly highlights the effectiveness of the strategy in mitigating the impact of class imbalance. Finally, the effect of increasing the initial batch size is similar to that discussed above. Fig. \ref{fig:logg_comm} shows scores of QBC sampling applied to surface gravity with increasing initial batch sizes. Unlike the case of $T_{\text{eff}}$, there is no noticeable improvement in performance compared to uncertainty sampling strategies shown in Fig. \ref{fig:logg_non_comm}. Again, this might be due to the use of committee members without prior hyperparameter tuning, which would require an initial labelled set for performing grid searches. It is worth noting that the vote-entropy strategy outperforms all others. The difference is particularly significant for MCC and sensitivity.
	
\begin{figure*}
		\centering
		\includegraphics[width=\textwidth]{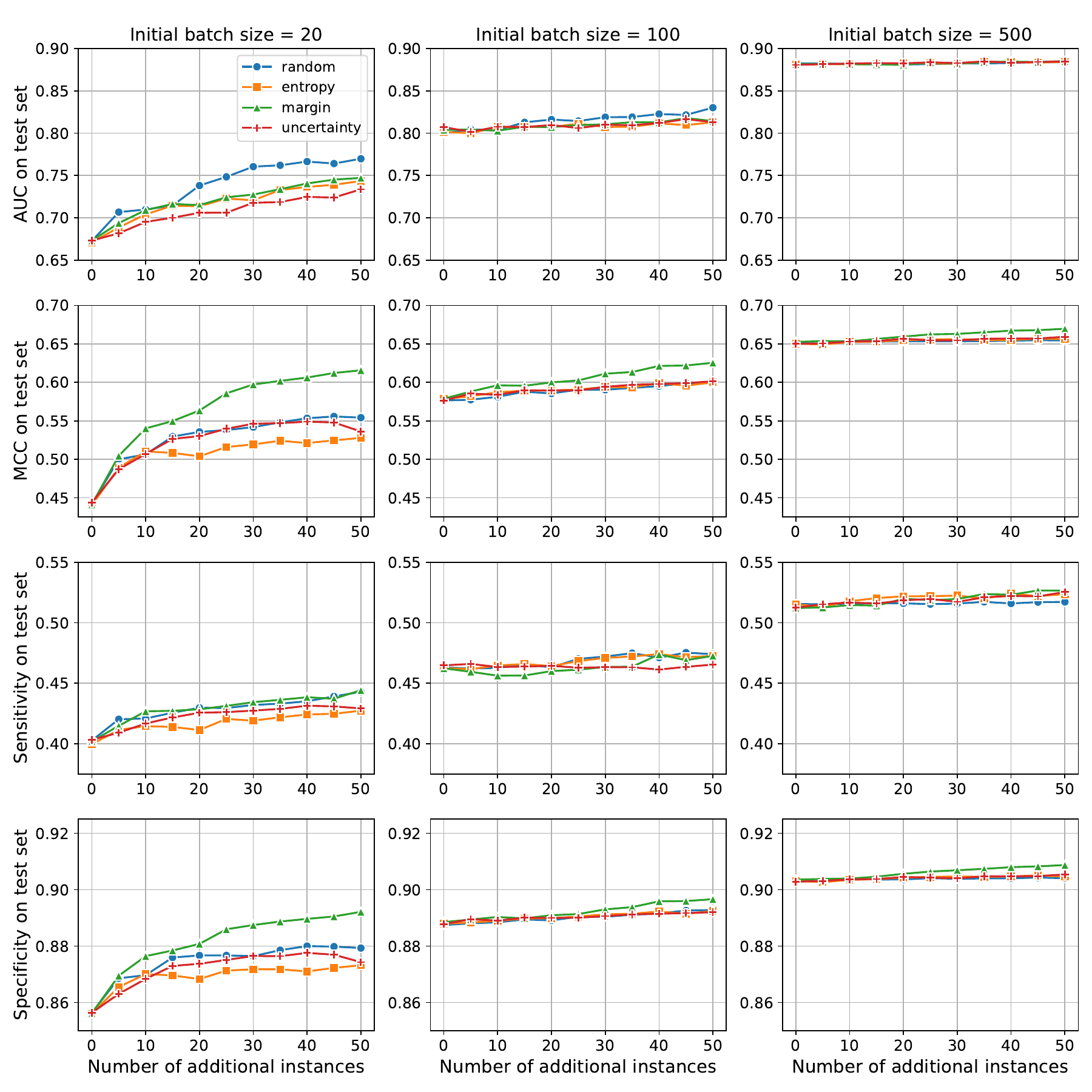}
		\caption{Performance scores for uncertainty sampling strategies along with random sampling for different initial batch sizes applied to [Fe/H].}
		\label{fig:feh_non_comm}
	\end{figure*}

	In Fig. \ref{fig:feh_non_comm}, we show the performance scores of uncertainty sampling strategies compared with random sampling when applied to iron metallicity with different initial batch sizes. Compared to the sensitivity scores when the model is applied to both $T_{\text{eff}}$ and $\log{g}$, we can see that the improvement with the increase in additional instances is much lower for most strategies in the case of [Fe/H]. This could be due to the existence of chemically peculiar (CP) stars in the minority classes of the testing set. CP stars mainly belong to spectral classes A and B \citep{Ghazaryan2018,Ghazaryan2019}; and their existence in the testing set will necessitate a higher number of training instance before the model can start to correctly identify rare classes. This is evident when we look at the improvement of sensitivity scores when we start with a batch size of 500 instances. It is worth noting that this impact will not be so pronounced if we choose the `micro' instead of `macro' values for the metrics (see Section \ref{subsec:metrics}). The figure also shows that uncertainty-margin sampling still outperforms all others in MCC, sensitivity, and specificity, even with the increase of $n_{\text{init}}$.
	
\begin{figure*}
		\centering
		\includegraphics[width=\textwidth]{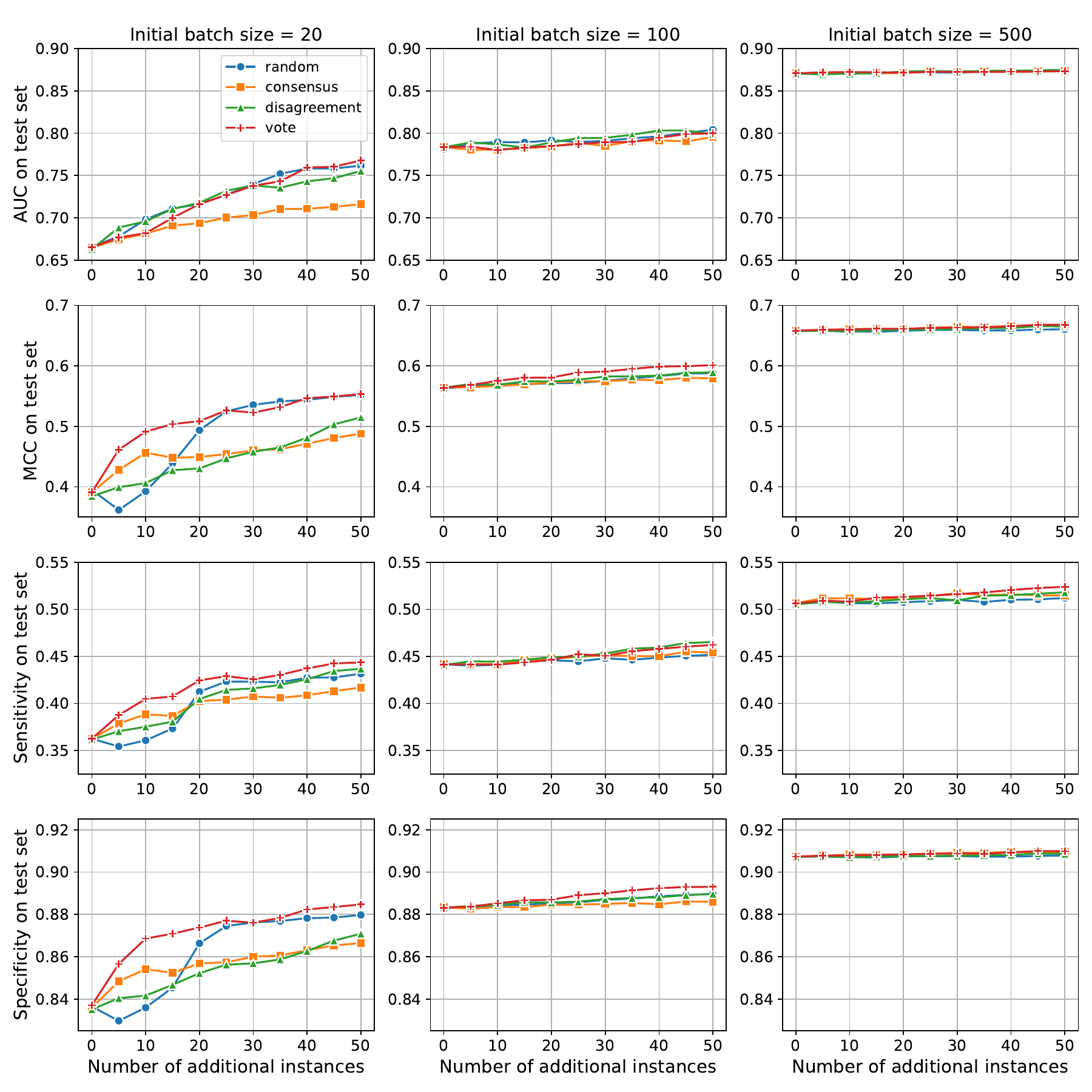}
		\caption{Performance scores for QBC disagreement sampling strategies along with random sampling for different initial batch sizes applied to [Fe/H].}
		\label{fig:feh_comm}
	\end{figure*}

	The last plots for the first part of this study are shown in Fig. \ref{fig:feh_comm}; namely, the performance scores of QBC sampling strategies applied to iron metallicity with increasing initial batch sizes. When compared to Fig. \ref{fig:feh_non_comm}, we can see that QBC does not offer any improvement upon single-learner uncertainty sampling in any metric, regardless of the additional computational cost of training a QBC model. Some models show erratic behaviour at lower numbers of additional instances with $n_{\text{init}} = 20$. This could be due to the KNN member of the committee, which requires more class population to stabilize (particularly in case of existence of CP stars). This is particularly evident for random sampling because it is less likely to populate neighbourhoods of rare classes with fewer instances. It is worth mentioning that the vote-entropy strategy still outperforms all others across all metrics and initial batch sizes.
	
	Taking a closer look at Fig. \ref{fig:teff_comm} and Fig. \ref{fig:logg_comm} together, we can see an `elbow' feature in the first column ($n_{\text{init}} = 20$) across all metrics at 15 additional instances, which is equivalent to a total training set size of 35 instances. However, this feature does not appear in Fig. \ref{fig:teff_non_comm} or Fig. \ref{fig:logg_non_comm}, indicating that it can be attributed to the use of QBC. The most likely reason is that each member of the committee requires the feature space to be populated in a different way in order to improve performance. This translates to the necessity of a higher number of training instances to improve the collective performance of the committee. This is even corroborated by the fact that we can not see a similar feature in Fig. \ref{fig:feh_comm} because the number of iron metallicity classes is lower (4 compared to 7 and 6 for $T_{\text{eff}}$ and $\log{g}$, respectively).
	
	Finally, Table \ref{tab:AL_stats} summarizes the statistics describing the impact of the different AL strategies on the classification performance, each compared to random sampling. For each of the metrics listed in Section \ref{subsec:metrics}, we calculated the $p$-value of the Wilcoxon's signed-rank test and Cohen's $d$ impact index. A $p$-value of less than 0.05 indicates a statistically significant impact, and a positive Cohen's $d$ shows superiority of the active learning strategy over random sampling and vice versa. These results were calculated by aggregating across all runs and initial batch sizes for each parameter-strategy-metric combination. In addition to validating our findings based on Figures \ref{fig:teff_non_comm}--\ref{fig:feh_comm}, the following points are worth noting:
	\begin{itemize}
		\item The classification margin strategy results in highly significant ($p\leq8.51\times10^{-4}$) improvements in MCC and sensitivity across all parameters.
		\item The vote-entropy strategy shows highly significant ($p\leq1.21\times10^{-3}$) improvements across all metrics for both $T_{\text{eff}}$ and $\log{g}$ classification.
		\item All AL uncertainty strategies show higher sensitivity results for both $T_{\text{eff}}$ and $\log{g}$ classification.
	\end{itemize}
	
	\begin{table*}
		\caption{The statistical impact of different AL strategies on classification metrics compared to random sampling, where the $p$-value is calculated for the Wilcoxon's signed-rank test.}
		\label{tab:AL_stats}
		\centering
		\footnotesize
		\begin{tabular}{ccccccccc}
			\hline\hline
			\multirow{2}{*}{Strategy}	& \multicolumn{2}{c}{AUC}				& \multicolumn{2}{c}{MCC}				& \multicolumn{2}{c}{Sensitivity}	& \multicolumn{2}{c}{Specificity}\\
			& $p$-value				& Cohen's $d$	& $p$-value				& Cohen's $d$	& $p$-value			& Cohen's $d$	& $p$-value				& Cohen's $d$\\
			\hline
			\multicolumn{9}{c}{$T_{\text{eff}}$: Non-committee strategies}\\
			\hline
			Entropy		& $2\times10^{-6}$		& 2.09			& $2.15\times10^{-2}$	& $-0.53$		& $2\times10^{-6}$	& 1.90			& $1.21\times10^{-3}$	& $-0.75$\\
			Margin		& $5.86\times10^{-4}$	& 0.81			& $3.6\times10^{-5}$	& $1.19$		& $4\times10^{-6}$	& 1.49			& $8.2\times10^{-5}$	& $1.11$\\
			Uncertainty	& $2\times10^{-6}$		& 2.27			& $6.22\times10^{-1}$	& $-0.17$		& $2\times10^{-6}$	& 2.22			& $5.32\times10^{2}$	& $-0.46$\\
			\hline
			\multicolumn{9}{c}{$T_{\text{eff}}$: Committee strategies}\\
			\hline
			Consensus	& $2\times10^{-6}$		& 1.67			& $8.51\times10^{-4}$	& $-0.87$		& $2\times10^{-6}$		& 2.02			& $3.22\times10^{-4}$	& $-0.91$\\
			Disagreement& $6\times10^{-6}$		& 1.33			& $7.30\times10^{-3}$	& $-0.52$		& $3.22\times10^{-4}$	& 0.92			& $7.30\times10^{-3}$	& $-0.59$\\
			Vote		& $4\times10^{-6}$		& 1.52			& $1.05\times10^{-4}$	& $0.99$		& $2\times10^{-6}$		& 2.22			& $3.6\times10^{-5}$	& $1.04$\\
			\hline
			\multicolumn{9}{c}{$\log{g}$: Non-committee strategies}\\
			\hline
			Entropy		& $3.62\times10^{-2}$ & $-0.53$ & $6.48\times10^{-1}$ & $-0.11$ & $1.89\times10^{-1}$ & 0.34 & $6.74\times10^{-1}$ & $-0.24$\\
			Margin		& $2.15\times10^{-2}$ & $0.56$  & $3.95\times10^{-4}$ & $1.18$  & $1.05\times10^{-4}$ & 1.16 & $3.15\times10^{-3}$ & $0.85$\\
			Uncertainty	& $7.56\times10^{-1}$ & $-0.13$ & $2.02\times10^{-1}$ & $0.26$  & $1.53\times10^{-2}$ & 0.67 & $3.88\times10^{-1}$ & $0.10$\\
			\hline
			\multicolumn{9}{c}{$\log{g}$: Committee strategies}\\
			\hline
			Consensus	& $2.31\times10^{-1}$ & $-0.30$ & $2.02\times10^{-1}$ & $-0.45$ & $2.45\times10^{-1}$ & $0.22$  & $3.62\times10^{-2}$ & $-0.65$\\
			Disagreement& $3.49\times10^{-1}$ & $-0.11$ & $1.89\times10^{-1}$ & $-0.25$ & $6.74\times10^{-1}$ & $-0.03$ & $2.15\times10^{-2}$ & $-0.488$\\
			Vote		& $4.8\times10^{-5}$  & $1.36$  & $4.8\times10^{-5}$  & $1.41$  & $4\times10^{-6}$    & $1.73$  & $1.21\times10^{-3}$ & $1.04$\\
			\hline
			\multicolumn{9}{c}{[Fe/H]: Non-committee strategies}\\
			\hline
			Entropy		& $1.72\times10^{-2}$ & $-0.50$ & $5.83\times10^{-2}$ & $-0.61$ & $5.32\times10^{-2}$ & $-0.31$ & $2.15\times10^{-2}$ & $-0.75$\\
			Margin		& $2.45\times10^{-1}$ & $-0.25$ & $2\times10^{-6}$    & $2.28$  & $8.51\times10^{-4}$ & $0.58$  & $4\times10^{-6}$    & $1.61$\\
			Uncertainty	& $7.30\times10^{-3}$ & $-0.46$ & $1.33\times10^{-1}$ & $-0.49$ & $8.70\times10^{-2}$ & $-0.26$ & $3.62\times10^{-2}$ & $-0.63$\\
			\hline
			\multicolumn{9}{c}{[Fe/H]: Committee strategies}\\
			\hline
			Consensus	& $7.30\times10^{-3}$ & $-0.52$ & $2.10\times10^{-4}$ & $-1.09$ & $8.26\times10^{-2}$ & $-0.42$ & $2.61\times10^{-4}$ & $-0.94$\\
			Disagreement& $4.09\times10^{-1}$ & $-0.14$ & $3.15\times10^{-3}$ & $-0.75$ & $7.01\times10^{-1}$ & $-0.15$ & $1.53\times10^{-2}$ & $-0.62$\\
			Vote		& $2.94\times10^{-1}$ & $-0.16$ & $9.27\times10^{-1}$ & $0.05$  & $2.61\times10^{-1}$ & $0.18$  & $2.94\times10^{-1}$ & $0.17$\\
			\hline
		\end{tabular}
	\end{table*}
	
\begin{figure}
		\centering
		\includegraphics[width=\columnwidth]{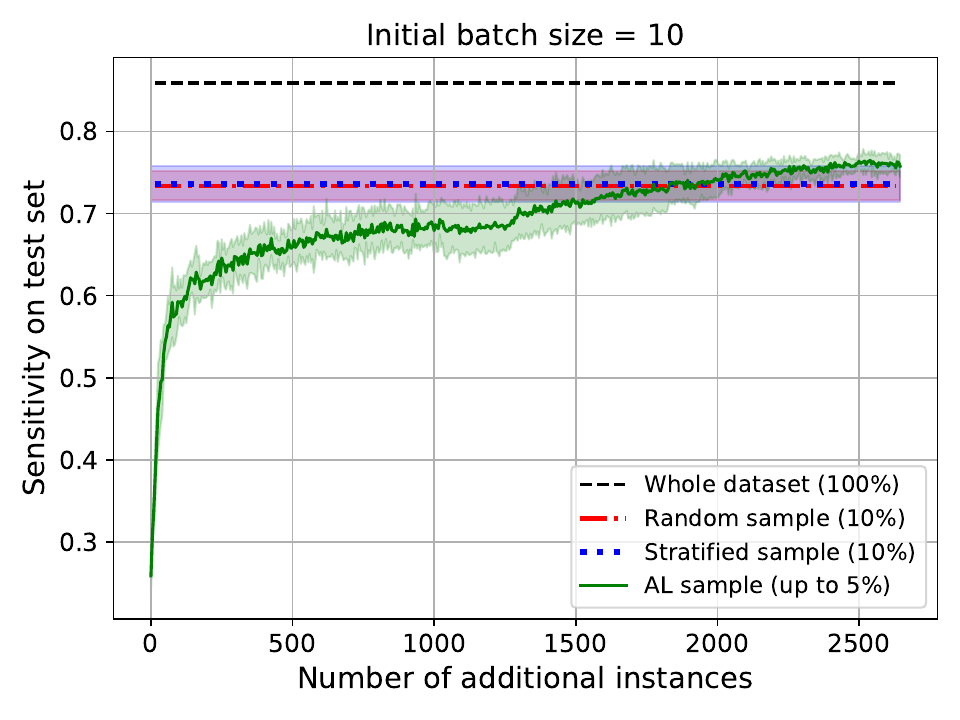}
		\caption{Progress of the sensitivity score on the test set with the increase in the number of additional instances queried, using an RF model trained on a sample selected by the uncertainty-margin strategy with an initial batch size of 10 instances applied to $T_{\text{eff}}$, with several baselines drawn for reference showing the score using the whole training set pool, a random sample of 10\%, and a stratified sample of 10\% as well.}
		\label{fig:sen_data_perc}
	\end{figure}

	In the second part of this study, we used an RF model along with the uncertainty-margin sampling strategy to track the progress of the sensitivity score on the test set with the increase in the number of instances queried by the AL algorithm; the results -- when applied to effective temperature -- are shown in \mbox{Fig. \ref{fig:sen_data_perc}}. We also included three baselines to reference for comparison, corresponding to RF models trained on three different sets: the whole training pool (100\%), a random sample of 10\%, and a stratified sample of 10\% as well. For the AL strategy, we use only 10 instances as a random initial batch in this experiment. Based on these results, on one hand, it seems that stratification does not add any improvement over random sampling. This demonstrates that a sampling strategy more effective than stratification is needed to achieve higher performance scores with fewer training instances and less computational cost, even disregarding the fact that stratified sampling requires the entire data pool to be labelled prior to instance selection. On the other hand, it is clear that the AL approach outperforms both samples with only half the training-set size. We can also see that the variance in the AL approach sensitivity starts to increase after the first score jump at around 100 instances. This is because the feature space of the training sample is widening but has not yet accumulated enough instances to cover the finer details of each class. However, when the training sample reaches a size of around $1,500$ instances, we can see the variance starting to significantly decrease. In spite of this, the sensitivity score of the uncertainty-margin algorithm is still trending upward to the end of the curve, which shows that further improvement can be safely expected when we increase the number of sampled instances even further, when more computational resources are available.
	
	Other AL algorithms and ML models can be utilized to perform the above experiments as regression problems rather than classification ones, if we wish to estimate the parameters associated with each spectrum instead of classifying it. At the moment, our computational resources do not permit us to perform the study in this format. However, this work is meant as a proof-of-concept for the effectiveness of the AL approach, in general,  in curating training sets for ML models.
	
	\section{Conclusions}
	\label{sec:conclusions}
	The results shown in this paper demonstrate the effectiveness of AL in curating training sets for supervised ML models with the objective of achieving the best possible stellar spectral classification performance while reducing labelling costs in terms of time and expertise. Compared to classical classification approaches, there are several interesting conclusions. They are as follows:
	\begin{enumerate}
		\item AL algorithms significantly improve the performance of stellar spectral classification compared to random or stratified sampling methods, by iteratively selecting the most informative instances to annotate.
		\item AL reduces the size of the labelled training set required for achieving the same performance as random sampling, making it more cost-effective and efficient.
		\item AL algorithms are more robust against class imbalance, which is often the case in stellar spectra datasets, consequently ensuring that rarer stellar classes would be represented adequately and, thus, classified correctly.
		\item AL sampling strategies are scalable and can be practically used on large datasets, indicating that it can be integrated into stellar survey data processing pipelines.
		\item Models trained on samples curated using AL methods exhibit better generalization with fewer instances, which is evident when evaluated on unseen testing data, making them more reliable in real-survey applications.
	\end{enumerate}
	
	Therefore, the AL approach for automating stellar spectra data curation and classification is feasible, accurate, and cost-effective. Based on the findings of this study, we recommend the integration of AL algorithms in citizen science projects to accelerate the annotation process even further. They can also be used in automated astronomical surveys to optimize the selection of spectra for follow-up observations and analysis. Future work will be conducted to:
	\begin{enumerate}
		\item adapt the approach used here for multi-label classification in order to further minimize the amount of training data needed;
		\item investigate the percentage of data required to achieve the same performance scores obtained by using the entire dataset;
		\item merge data from different surveys to use in curating a comprehensive training sample using AL, and make it publicly available to use for automated stellar classification in future surveys;
		\item evaluate AL algorithms available for regression problems, in order to leverage them in curating pipelines for estimation of stellar atmospheric parameters.
	\end{enumerate}
	All of the above is contingent on increasing the computational resources currently available.
	
	\begin{acknowledgements}
		To prepare the code required for performing this work, we used each of the following open-source Python \citep{van1995python} libraries: \texttt{NumPy} \citep{Harris2020}, \texttt{pandas} \citep{pandas2024,McKinney2010}, \texttt{Matplotlib} \citep{Hunter2007}, \texttt{Scikit-learn} \citep{Pedregosa2011}, \texttt{Imbalanced-learn} \citep{Lemaitre2017}, \texttt{Astropy} \citep{Robitaille2013,PriceWhelan2018,Collaboration2022}, and \texttt{modAL} \citep{Danka2018}.
		
		In this work, we have used the SDSS database extensively. Funding for the Sloan Digital Sky Survey IV has been provided by the Alfred P. Sloan Foundation, the U.S. Department of Energy Office of Science, and the Participating Institutions. 
		
		SDSS-IV acknowledges support and resources from the Center for High Performance Computing  at the University of Utah. The SDSS website is \url{www.sdss4.org}.
		
		SDSS-IV is managed by the Astrophysical Research Consortium for the Participating Institutions of the SDSS Collaboration including the Brazilian Participation Group, the Carnegie Institution for Science, Carnegie Mellon University, Center for Astrophysics | Harvard \& Smithsonian, the Chilean Participation Group, the French Participation Group, Instituto de Astrof\'isica de Canarias, The Johns Hopkins University, Kavli Institute for the Physics and Mathematics of the Universe (IPMU) / University of Tokyo, the Korean Participation Group, Lawrence Berkeley National Laboratory, Leibniz Institut f\"ur Astrophysik Potsdam (AIP),  Max-Planck-Institut f\"ur Astronomie (MPIA Heidelberg), Max-Planck-Institut f\"ur Astrophysik (MPA Garching), Max-Planck-Institut f\"ur Extraterrestrische Physik (MPE), National Astronomical Observatories of China, New Mexico State University, New York University, University of Notre Dame, Observat\'ario Nacional / MCTI, The Ohio State University, Pennsylvania State University, Shanghai Astronomical Observatory, United Kingdom Participation Group, Universidad Nacional Aut\'onoma de M\'exico, University of Arizona, University of Colorado Boulder, University of Oxford, University of Portsmouth, University of Utah, University of Virginia, University of Washington, University of Wisconsin, Vanderbilt University, and Yale University.
		
		We are grateful to the anonymous referee for their insightful comments and suggestions, which have significantly improved the quality and clarity of this manuscript.
	\end{acknowledgements}
	
	%
	\bibliographystyle{aa} 
	\bibliography{refs} 
	%
	
\end{document}